\documentclass[arps,prd,twocolumn,showpacs]{revtex4}
\usepackage{graphicx}
\usepackage{amsmath}
\usepackage{amsfonts}
\usepackage{amssymb}
\usepackage{color}
\usepackage{epsf}

\newcommand{\be}{\begin{equation}}         
\newcommand{\ee}{\end{equation}}
\newcommand{\ba}{\begin{eqnarray}}
\newcommand{\ea}{\end{eqnarray}}
\newcommand{\nn}{\nonumber}

\newcommand\lsim{\mathrel{\rlap{\lower4pt\hbox{\hskip1pt$\sim$}}
        \raise1pt\hbox{$<$}}}
\newcommand\gsim{\mathrel{\rlap{\lower4pt\hbox{\hskip1pt$\sim$}}
        \raise1pt\hbox{$>$}}}

\def\x{{\bf x}}

\begin{document}

\title{ Super cosmic variance from mode coupling: A worked example}
\author{ Marilena LoVerde}
\affiliation{Enrico Fermi Institute, Kavli Institute for Cosmological Physics, Department of Astronomy and Astrophysics, University of Chicago, Illinois, 60637, USA \\ marilena@uchicago.edu}
\begin{abstract}

If the entire post-inflationary patch is large compared to our Hubble volume even a small level of non-Gaussianity can cause statistics of the primordial curvature field in our Hubble volume to be biased by mode coupling.
We explicitly compute the variation of locally measured statistics of the primordial curvature $\zeta$ from non-Gaussian mode coupling within a specific inflationary scenario: the curvaton model with a quadratic curvaton potential. This ``super cosmic variance" is calculated in two ways: (i) as a super observer who has access to the curvature perturbation field across the entire post-inflationary patch and therefore sees local statistics as biased by mode coupling and (ii) as a local observer who sees the statistics of $\zeta$ determined by the local values of quantities in their Hubble patch. The two calculations agree and show that in the quadratic curvaton model patch-to-patch differences in statistics of $\zeta$ can be interpreted entirely as a shift in the value of the curvaton field at freeze out. Applying the same arguments to single-field slow-roll inflation gives a simple picture of how non-Gaussian mode coupling between the curvature perturbations on very different physical scales must vanish in the attractor limit. 
\end{abstract}
\maketitle

\section{Introduction}
\label{sec:intro}
Determining the physical mechanism responsible for seeding cosmic structure, for instance inflation, is one of the most exciting possibilities in cosmology. Our primary data from this era is the primordial inhomogeneities in curvature that give rise to anisotropies in the cosmic microwave background and seed cosmic structures such as galaxies \cite{Guth:1980zm,Mukhanov:1981xt,Hawking:1982cz,Starobinsky:1982ee, Guth:1982ec,Bardeen:1983qw}. The statistics of the primordial curvature $\zeta$, i.e. the power spectrum, bispectrum and trispectrum, can provide powerful information about inflationary models and alternative scenarios (see, for instance \cite{Baumann:2008aq}). Cosmic microwave background data puts stringent limits on deviations from Gaussian statistics for $\zeta$, consistent with the predictions from the simplest single-field, slow-roll inflationary models \cite{Ade:2013zuv,Ade:2013uln,Ade:2013ydc}. 

Observational constraints on the statistics of $\zeta$ are limited to measurements from within our Hubble volume but our Hubble volume may not be representative of the entire post-inflationary patch. For instance, local-type primordial non-Gaussianity induces correlations between small-scale fluctuations in $\zeta$, measurable on sub-Hubble scales, and long-wavelength fluctuations in $\zeta$ that are unmeasurable from within our Hubble volume. If the entire post-inflationary patch is large in comparison with our Hubble volume, the variance of these long-wavelength fluctuations in $\zeta$ is large and observations of local statistics can be biased by coupling between modes on sub-Hubble and super-Hubble scales \cite{Fan:1995aq,Boubekeur:2005fj,Gordon:2005ai,Schmidt:2012ky,Nelson:2012sb,Nurmi:2013xv,LoVerde:2013xka,Bramante:2013moa,Byrnes:2013qjy}. In this paper we refer to the variation in observational quantities among different Hubble-sized patches as ``super cosmic variance" and changes to local quantities specifically from coupling between super and sub-Hubble scales as ``mode coupling bias."

In \cite{Nelson:2012sb,Nurmi:2013xv,LoVerde:2013xka} the mode-coupling bias is calculated from a statistical perspective. More precisely, it is assumed that the field $\zeta$ is defined across the entire post-inflationary patch and changes to the local statistics of $\zeta$ arise from coupling between short- and long- wavelength modes. We refer to this mode-coupling calculation as from the perspective of a super observer, because it relies on knowing $\zeta$ across the entire post-inflationary patch.  The purpose of this paper is to provide an alternative calculation of mode-coupling bias from the perspective of a sub-Hubble observer. In this second calculation, sub-Hubble quantities depend on the local background (e.g. the local expansion rate and/or energy density in some component) which vary among Hubble-sized patches. To make things concrete we work with the curvaton scenario  \cite{Mollerach:1989hu,Linde:1996gt,Lyth:2001nq}, a particular two-field inflationary model that generates local-type non-Gaussianity.  As we shall see, the mode-coupling calculation agrees with the calculation of the local curvature perturbation  in a background with parameter values shifted according to the amplitude of long-wavelength modes. Agreement between the two calculations is expected and coincides with the physical picture that local-type non-Gaussianity arises when the amplitude of curvature perturbations that are generated depends on the background. In the curvaton scenario, the background dependence arises from the dependence of $\zeta$ on the freeze-out value of the curvaton. In the curvaton scenario, it is easy to see how local statistics may appear nearly Gaussian in a Hubble-sized patch, even if $\zeta$ is strongly non-Gaussian across the entire post-inflationary patch: nearly Gaussian patches are merely regions where the curvaton field was displaced far from the potential minimum during inflation.

The bias from mode coupling discussed in this paper is just one aspect of the more general question of how to relate the values of quantities measured in our Hubble volume to those in the rest of the universe. For instance, even for a finite post-inflationary patch one might want to incorporate anthropic considerations to relate predictions from inflation to predictions for local observations (for example, \cite{Salem:2012ve}). On the other hand, allowing for the possibility of eternal inflation \cite{Vilenkin:1983xq,Linde:1986fc} complicates things further as predictions from inflation vary dramatically depending on how one assigns probabilities to different regions of spacetime (see for instance, \cite{Freivogel:2011eg,Salem:2011qz} and references therein). The mode-coupling bias discussed here and in \cite{Nelson:2012sb,Nurmi:2013xv,LoVerde:2013xka,Bramante:2013moa} is related to these issues, but the author considers it a simpler problem that can be studied independently. More precisely, changes to local statistics from mode coupling can arise in a finite, non-eternally inflating patch. And, though one can include an anthropic prior on the probability to observe particular values of quantities in our Hubble patch, the problem of choosing that prior can be considered separately from the volume-weighted probability that particular values of those local quantities exist. 

The rest of this paper is organized as follows.  Section \ref{sec:superobserver} reviews the calculation from \cite{LoVerde:2013xka} relating local and global statistics from the perspective of the super observer (see also, \cite{Nelson:2012sb,Nurmi:2013xv}). The calculation in \S\ref{sec:superobserver} gives a general expression for the mapping from global to local quantities without reference to a particular inflationary model. In \S \ref{sec:curvaton}, we calculate curvature perturbations in the curvaton model and compare the prediction from  \S \ref{sec:superobserver} to what is seen by a local observer with background parameters shifted by long-wavelength curvaton perturbations. In \S \ref{sec:inflaton} we repeat the calculation of the dependence of local $\zeta$ on long wavelength perturbations in single-field slow-roll inflation. For single-field slow-roll inflation on an attractor solution the local curvature power spectrum does not vary with long-wavelength perturbations in the inflaton.

\section{Curvature perturbations from the perspective of the super observer}
\label{sec:superobserver}
Before proceeding we review definitions of the locally and globally defined curvature perturbation and the expressions for mode coupling in the presence of weak, local-type primordial non-Gaussianity. The discussion here follows \cite{LoVerde:2013xka} but see also \cite{Nelson:2012sb,Nurmi:2013xv,Bramante:2013moa}.

Following \cite{Lyth:2004gb}, let $\tilde{a}$ be the locally defined scale factor, and we factor $\tilde{a}$ into a spatially homogeneous piece and a perturbation, $ \tilde{a}=ae^{\zeta(\x)}$ where $\zeta$ is a perturbation with volume average equal to zero over a volume $V_L$ -- the region we have been referring to as the entire post-inflationary patch. In the uniform-density slicing the variable $\zeta$ corresponds to the conserved curvature perturbation on super-horizon scales. A local observer may define the curvature perturbation with respect to some smaller volume $V_S$ (for instance for our Hubble patch $V_S \sim H_0^{-3}$) as
\ba
\label{eq:volzetadefs}
\zeta_S(\x) &=& \ln \tilde{a}-\frac{1}{V_S}\int_{V_S}d^3\x \, \ln \tilde{a}\\
&=& \zeta(\x)-\zeta_L(\x)
\ea
where
\be
\zeta_L(x)\equiv \frac{1}{V_S}\int_{V_S}d^3\x \, \ln \tilde{a}-\frac{1}{V_L}\int_{V_L}d^3\x \, \ln \tilde{a}\,.
\ee
With these definitions a change in reference volume amounts to an additive shift in the perturbations by an amount $\zeta_L$. The long-wavelength mode $\zeta_L$ is unobservable and causes unknown, random shifts to locally measured parameters. Note that $\zeta_{L}$ has contributions from all modes with wavelengths longer than our horizon scale and is therefore boosted from the naive guess of $\sim 10^{-5}$ by roughly a factor of $N = \ln V_L/V_S$; for a more detailed discussion see \cite{LoVerde:2013xka}.

Consider a situation where the non-Gaussian statistics seen by the super observer (with access to all of $V_L$) can be written as a local non-linear transformation of a Gaussian field,  for instance
\ba
\label{eq:zetafNLgNL}
\zeta(\x)&=&\zeta_{G}(\x)+\frac{3}{5}f_{NL}\left(\zeta^2_G(\x)-\langle \zeta_G^2\rangle\right)\\
&+&\frac{9}{25}g_{NL}\left(\zeta^3_G(\x)-3\langle\zeta^2_G\rangle \zeta_G(\x)\right) \nn\\
&+& \frac{27}{125}h_{NL}\left(\zeta_G^4-6\langle\zeta_G^2\rangle\zeta_G^2(\x)+3\langle\zeta_G^2\rangle^2\right)+\dots\nn,
\ea
where $\langle \rangle $ indicates averages over $V_L$. Non-linearity in the variable $\zeta_{G}$ couples Fourier modes of $\zeta$ on different scales. Splitting the Gaussian field $\zeta_G$ in  Eq.~(\ref{eq:zetafNLgNL}) into short- and long- wavelength pieces, $\zeta_{G,S}$ and $ \zeta_{G,L}$, the curvature perturbation seen by a local observer (Eq.~(\ref{eq:volzetadefs})) is 
\ba
\left.\zeta\right|_{S} & = &\left(1+\frac{6}{5}f_{NL}\zeta_{G,L}\right)\zeta_{G,S} \\
&+& \frac{3}{5}\left( f_{NL}+\frac{9}{5}g_{NL}\zeta_{G,L}\right)\left(\zeta_{G,S}^2(\x) - \langle\zeta_{G,S}^2\rangle\right)\nn\\
&+& \dots + \mathcal{O}(\zeta_L^2)\,,\nn
\ea
where the $\dots$ indicate higher-order terms in $\zeta_{G,S}$. To lowest order in $\zeta_{G,L}$, the power spectrum and non-Gaussian parameters as measured locally in $V_S$ are
\be
\label{eq:Pmodecoupling}
\left.\Delta^2_{\zeta} \right|_{S}= \Delta^2_{\zeta}\left(1 + \frac{12}{5}f_{NL}\zeta_{G,L}\right)\,,
\ee
\be
\label{eq:fNLmodecoupling}
\left.f_{NL}\right|_{S} = f_{NL}+\frac{9}{5}g_{NL}\zeta_{G,L}-\frac{12}{5}f_{NL}^2\zeta_{G,L}\,,
\ee
\be
\label{eq:gNLmodecoupling}
\left.g_{NL}\right|_{S} = g_{NL}+\frac{12}{5}h_{NL}\zeta_{G,L}-\frac{18}{5}f_{NL}g_{NL}\zeta_{G,L} \,. 
\ee
Equations ~(\ref{eq:Pmodecoupling}) - (\ref{eq:gNLmodecoupling}) are general expressions for the locally measured statistics of the curvature perturbation in the presence of a long wavelength fluctuation $\zeta_{G,L}$. In the next few sections we will calculate perturbations as a function of the local background values of parameters, which may vary with $\zeta_{G,L}$, and show that the variation in $\Delta_\zeta^2$, $f_{NL}$, and $g_{NL}$ among regions with different $\zeta_{G,L}$ is precisely what is predicted by the mode coupling calculations given above.

\section{Curvature Perturbations in the curvaton model}
\label{sec:curvaton}
Let us now review the calculation of perturbations in the curvaton model. To keep things simple, assume that there is no coupling between the curvaton and inflaton and that the curvaton potential is exactly quadratic. Relaxing these assumptions leads to a wider range of possibilities for the statistics of $\zeta$ in the curvaton scenario (see for instance \cite{Linde:2005yw,Demozzi:2010aj}). 

The calculations here mostly follow the notation of  \cite{Lyth:2001nq,Sasaki:2006kq}, but we pay special attention to how averages are defined and how quantities are split between background and perturbations. Throughout we make the sudden-decay approximation, interpreted as saying the curvaton decays when the local Hubble rate is some fixed value $\Gamma$. In \S \ref{ssec:curvatondominates} we assume that the energy density of the universe is dominated by the curvaton at the time of curvaton decay, and in \S \ref{ssec:curvatonradiation} we allow contributions to the energy density from both curvaton and radiation. 

\subsection{Generation of curvaton perturbations during inflation}
We assume that there are no couplings between the inflaton and the curvaton and that the inflaton $\phi$ dominates the energy density during inflation. In this limit contributions from the curvaton $\sigma$ can be neglected in the Friedmann Equation: 
\be
\label{eq:rhophiHubble}
\rho_\sigma \ll \rho_\phi \,,\qquad  H^2 = \frac{\rho_{\phi}}{3M_{pl}^2}\,.
\ee
For a curvaton with a quadratic potential $V_{curv.}(\sigma) =\frac{1}{2}m^2\sigma^2$ the equation of motion for the spatially homogenous part is
\be
\label{eq:sigmabg}
\ddot\sigma + 3 H\dot\sigma + m^2\sigma = 0\,,
\ee
where $\dot{}$ is the derivative with respect to time.  On super horizon scales ($k/a \ll H$) the perturbations satisfy 
\be
\label{eq:sigmapert}
\ddot{\delta\sigma} + 3 H\dot{\delta\sigma} + m^2\delta\sigma = 0\,,
\ee
where the field perturbations are defined with respect to spatially flat hypersurfaces. For $m \ll H$, the spectrum of curvaton perturbations is 
\be
\label{eq:Deltasigma2}
\Delta_\sigma^2  =  \frac{H^2_{exit}}{(2\pi)^2}
\ee
where $H_{exit}$ is the value of the Hubble parameter when a mode with wavenumber $k$ exits the horizon. Note that since the inflaton dominates the energy density, $H_{exit}$ and therefore $\Delta_\sigma^2$ are completely determined by $\phi$. In particular, the variance of curvaton perturbations $\Delta_\sigma^2$ is not affected by long-wavelength fluctuations in $\sigma$. 

\subsection{Curvature perturbations from the curvaton I: curvaton dominates the energy density at decay}
\label{ssec:curvatondominates}
Suppose that at the end of inflation the curvaton freezes out at some value $\sigma_*$. We assume that there is no evolution of the curvaton field on super horizon scales so that $\sigma_*$ is also the initial amplitude of the field when the curvaton begins to oscillate about the potential minimum (see \cite{Sasaki:2006kq} for possible changes to for non-quadratic curvaton potentials). During this period $\rho_\sigma \propto a^{-3}$. 

 Curvature perturbations are generated when the curvaton decays. In the sudden-decay approximation, curvaton decay happens when the local density $\rho_\sigma =  3\Gamma^2M_{pl}^2$ where $\Gamma$ is the decay rate. Regions with different values of $\delta\sigma$, say $\delta\sigma_1$ and $\delta\sigma_2$, will have differences in the curvature perturbation $\zeta_1$ and $\zeta_2$ related through  
\be
\rho_\sigma(a)\left(1+\frac{\delta\sigma_1}{\sigma_*}\right)^2e^{-3\zeta_1}=\rho_\sigma(a)\left(1+\frac{\delta\sigma_2}{\sigma_*}\right)^2e^{-3\zeta_2}\,.
\ee
The difference in the expansion between two regions with different $\delta\sigma$ values is, 
\ba
\label{eq:curvnonlinear}
\zeta_1-\zeta_2 &=& \frac{2}{3}\ln\left(\frac{1+\frac{\delta\sigma_1}{\sigma_*}}{1+\frac{\delta\sigma_2}{\sigma_*}}\right)\\
&=& \frac{2}{3}\frac{\delta\sigma_1-\delta\sigma_2}{\sigma_*}-\frac{1}{3}\frac{\delta\sigma_1^2-\delta\sigma_2^2}{\sigma_*^2}\nn\\
&+&\frac{2}{9}\frac{\delta\sigma_1^3-\delta\sigma_2^3}{\sigma_*^3} - \frac{1}{6}\frac{\delta\sigma_1^4 - \delta\sigma_2^4}{\sigma_*^4}+ \dots\nn
\ea
Defining
\be
\label{eq:zetacurv}
\zeta_G(\x)=\frac{2}{3}\frac{\delta\sigma}{\sigma_*}(\x)
\ee
gives
\be
\label{eq:fNLgNLhNLcurv}
f_{NL} = -\frac{5}{4} \, , \quad g_{NL} = \frac{25}{12}\, ,\,{\rm and} \quad h_{NL} = -\frac{125}{32}\,.
\ee
Plugging these values into Eqs.~(\ref{eq:Pmodecoupling}) - (\ref{eq:gNLmodecoupling}), one predicts that an observer on a long-wavelength fluctuation $\zeta_{G,L}$ will see the power spectrum changed by a factor $(1 - 3\zeta_{G,L})$ but the values of $f_{NL}$ and $g_{NL}$ are {\em unchanged},

\ba
\label{eq:Pshiftcurv}
\left. \Delta^2_{\zeta}\right|_S &=& \Delta^2_\zeta(1 - 3\zeta_{G,L})\, \\
\label{eq:fNLshiftcurv}
\left.f_{NL} \right|_S &= & f_{NL}\,\\
\label{eq:gNLshiftcurv}
\left.g_{NL}\right|_S  &= & g_{NL}\,.
\ea
Note however that while the numerical values of $f_{NL}$ and $g_{NL}$ are unchanged by a long-wavelength mode, the level of non-Gaussianity as characterized by the dimensionless cumulants $B_{\zeta\zeta\zeta}/(P_\zeta)^{3/2} \sim f_{NL} \sqrt{\Delta_{\zeta}^2}$, $T_{\zeta\zeta\zeta\zeta} \sim (f_{NL}^2 + g_{NL})\Delta_\zeta^2$, does vary with $\zeta_{G,L}$. 


\subsubsection{Perturbations calculated by a local observer on a long-wavelength fluctuation $\delta\sigma_L$}

Now we calculate the small-scale statistics of the curvature in a region with a long-wavelength fluctuation in $\sigma$. Since it is the inflaton that determines the curvaton fluctuations $\delta\sigma$, the sole effect of a long-wavelength perturbation in the curvaton is to change the reference value of $\sigma$: $\sigma_*\rightarrow \sigma_*+\delta\sigma_L$ but the total $\frac{\rho_\sigma(a)}{3M_{pl}^2} = H_{decay}^2 = \Gamma^2$ remains fixed.
For regions with different values of $\delta\sigma$, but having a common $\delta\sigma_L$, this gives
\ba
\left(1+\frac{\delta\sigma_1}{\sigma_*(1+\frac{\delta\sigma_L}{\sigma_*})}\right)^2e^{-3\zeta_1}\qquad\qquad\qquad\\
=\left(1+\frac{\delta\sigma_2}{\sigma_*(1+\frac{\delta\sigma_L}{\sigma_*})}\right)^2e^{-3\zeta_2}\,,\nn
\ea
giving
\be
\label{eq:zetaScurv}
\zeta(\x)=\zeta_G(\x)-\frac{3}{5}\frac{5}{4}\left(\zeta_G(\x)^2-\langle\zeta_G^2\rangle\right)+\frac{9}{25}\frac{25}{12}\zeta_G^3(\x)+\dots\,,
\ee
where in this case 
\be
\zeta_G(\x)=\frac{2}{3}\frac{\delta\sigma}{\sigma_*}(\x)\left(1-\frac{\delta\sigma_L}{\sigma_*}\right)\,.
\ee
One sees that the coefficients in the expansion Eq.~(\ref{eq:zetaScurv}), giving the local values of $f_{NL}$ and $g_{NL}$,  are unchanged from Eq.~(\ref{eq:fNLgNLhNLcurv}) but the amplitude of fluctuations {\em is } changed from Eq.~(\ref{eq:zetacurv}) by a factor $1-\delta\sigma_L/\sigma_*$. The shift in amplitude changes the small-scale power spectrum by a factor $1-2\delta\sigma_L/\sigma_*$ or
\ba
\left.\Delta_\zeta^2\right|_S &=& \Delta_\zeta^2\left(1-3\zeta_{G,L}\right)\,.
\ea 
This expression for the shift in power, and the non-shifts of the local values of $f_{NL}$ and $g_{NL}$ are precisely what was predicted in Eqs.~(\ref{eq:Pshiftcurv}) -  Eq.~(\ref{eq:gNLshiftcurv}) in \S\ref{ssec:curvatondominates}. Note that it can already be seen in Eq.~(\ref{eq:curvnonlinear}) that $f_{NL}$ and $g_{NL}$ will not vary with $\delta\sigma_L$: a long-wavelength mode changes $\delta\sigma/\sigma_*$, not the numerical coefficients of $\delta\sigma/\sigma_*$ in the Taylor expansion of Eq.~(\ref{eq:curvnonlinear}).

\subsection{Generation of curvature perturbations from the curvaton II: curvaton and radiation present when curvaton decays}
\label{ssec:curvatonradiation}
Now we consider the case where both the curvaton and radiation are present when the curvaton decays and assume that there are no perturbations to the radiation energy density. The curvature perturbations are determined by
\ba
\rho_\sigma(a)\left(1+\frac{\delta\sigma_1}{\sigma_*}\right)^2e^{-3\zeta_1}+\rho_r(a)e^{-4\zeta_1}\\
=\rho_\sigma(a)\left(1+\frac{\delta\sigma_2}{\sigma_*}\right)^2e^{-3\zeta_2}+\rho_r(a)e^{-4\zeta_2}\nn\,.
\ea
We can expand in the small quantities $\delta\sigma$ and $\zeta$ and solve the above equation order by order to find the curvature perturbation $\zeta$. Defining
\be
r\equiv\frac{3\Omega_\sigma}{3\Omega_\sigma+4\Omega_r}
\ee
the expression for the curvature is
\ba
\label{eq:zetarsigma}
\zeta(\x)&=& \frac{2r}{3}\frac{\delta\sigma}{\sigma_*}(\x)\\
&+&\left(\frac{r}{3}-\frac{4r^2}{9}-\frac{2r^3}{9}\right)\left(\frac{\delta\sigma^2}{\sigma_*^2}(\x)-\langle\frac{\delta\sigma^2}{\sigma_*}\rangle\right)\nn\\
&+&\left(-\frac{4r^2}{9}+\frac{2r^3}{81}+\frac{40r^4}{81}+\frac{4r^5}{27}\right)\frac{\delta\sigma^3}{\sigma_*}(\x)\nn\\
&+&\dots\nn
\ea
Defining
\be
\zeta_G(\x)= \frac{2r}{3}\frac{\delta\sigma}{\sigma_*}(\x)\,,
\ee
one can write Eq.~(\ref{eq:zetarsigma}) as an expansion with the usual local form given in Eq.~(\ref{eq:zetafNLgNL}) with 
\ba
\label{eq:fNLcurvrad}
f_{NL} &=& \frac{5}{4r}-\frac{5}{3}-\frac{5r}{6} + \mathcal{O}(\langle\zeta_G^2\rangle) \\
\label{eq:gNLcurvrad}
g_{NL} &=& -\frac{25}{6r}+\frac{25}{108}+\frac{125r}{27}+\frac{25r^2}{18}\\
\label{eq:hNLcurvrad}
h_{NL} & =& -\frac{125}{48r^2} + \frac{2125}{288 r}+\frac{625}{36} - \frac{3125r}{324}\\
&& -\, \frac{4375r^2}{324} - \frac{625r^3}{216}\nn
\ea
in agreement with \cite{Sasaki:2006kq}.

\subsubsection{Perturbations calculated by a local observer on a long-wavelength fluctuation $\delta\sigma_L$}

We now solve for the curvature perturbation treating $\rho_\sigma$, $\rho_r$ as parameters that vary among patches with different values of $\delta\sigma_L$. 
One has,
\ba
\label{eq:LWandradnozetaL}
\rho_\sigma \left(1+\frac{\delta\sigma_L}{\sigma_*}\right)^2+\rho_r  \qquad\qquad \qquad \\
=\rho_\sigma \left(1+\frac{\delta\sigma_L}{\sigma_*}\right)^2\left(1+\frac{\delta\sigma}{\sigma_*(1+\frac{\delta\sigma_L}{\sigma_*})}\right)^2e^{-3\zeta}\nn\\
+\rho_re^{-4\zeta}\,.\nn
\ea
The local curvature perturbation is given by the same expression as before in Eq.~(\ref{eq:zetarsigma}),
\be
\zeta(\x)= \frac{2\left. r\right|_S}{3}\frac{\delta\sigma}{\sigma_*}(\x)\left(1-\frac{\delta\sigma_L}{\sigma_*}\right)\,,
\ee
but with a shift in $\sigma_*$ and the local value of $r$, here called $\left. r\right|_S$, shifted as well. The local value of $r$ is changed by both the long-wavelength contribution to $\rho_\sigma$ and the fact that the local decay time (in the region with $\delta\sigma_L$) is different, so the local values of $\rho_\sigma$ and $\rho_r$ used in the expressions for $\zeta$ will be different as well. The decay time (i.e. the time at which the surface of constant energy satisfies $(\rho_\sigma+\rho_r )/(3M_{pl}^2)= \Gamma^2$) is changed by an amount
\be
\delta t = -\frac{\delta\rho_\sigma}{\dot\rho_\sigma+\dot\rho_r}  = \frac{2\rho_\sigma}{H(3\rho_\sigma+4\rho_r)}\frac{\delta\sigma_L}{\sigma_*}\,.
\ee
The local values of $\rho_\sigma$ and $\rho_r$ at the time of decay are
\be
\left. \rho_\sigma\right|_S  =\rho_\sigma\left(1+(2-2r)\frac{\delta\sigma_L}{\sigma_*}\right)
\ee
and 
\be
\left. \rho_r\right|_S= \rho_r\left(1-\frac{8r}{3}\frac{\delta\sigma_L}{\sigma_*}\right)\,.
\ee
The local value of $r$ at curvaton decay is then
\be
\left. r\right|_S=r\left(1+\left(2-\frac{4r}{3}-\frac{2r^2}{3}\right)\frac{\delta\sigma_L}{\sigma_*}\right)
\ee
so the local variance of curvature perturbations is
\ba
\left. \Delta_\zeta^2\right|_S& =  &\Delta^2_\zeta\left(1+2\left(1-\frac{4r}{3}-\frac{2r^2}{3}\right)\frac{\delta\sigma_L}{\sigma_*}\right)\\
& =  &\Delta^2_\zeta\left(1+\frac{12}{5}f_{NL}\zeta_{G,L}\right)\nn
\ea
in agreement with the general expression for the shift in $\Delta_\zeta^2$ in terms of $f_{NL}$ and $\zeta_{G,L}$ given in Eq.~(\ref{eq:Pshiftcurv}). 

Shifting the values of $r$ in Eqs.~(\ref{eq:fNLcurvrad}) and (\ref{eq:gNLcurvrad}) gives the local values of $f_{NL}$ as
\ba
\left. f_{NL}\right|_S &=& f_{NL} - \left(\frac{15}{4r^2} - \frac{5}{2r} + \frac{5}{4} - \frac{5r}{3} - \frac{5r^2}{6}\right)\zeta_{G,L}\nn\\
&=&  f_{NL} + \left(\frac{9}{5}g_{NL} - \frac{12}{5}f_{NL}^2 \right)\zeta_{G,L}
\ea 
 and $g_{NL}$ as
\ba
\left.g_{NL} \right|_S &=& g_{NL}  +\nn\\
&& \left(\frac{25}{2r^2}- \frac{25}{3r}+ \frac{175}{18} -\frac{275r^2}{27} - \frac{25r^3}{9}\right)\zeta_{G,L} \nn\\
&=& g_{NL} + \left(\frac{12}{5}h_{NL} - \frac{18}{5}f_{NL}g_{NL}\right)\zeta_{G,L}
\ea
again in agreement with the predictions from mode-coupling equations Eq.~(\ref{eq:fNLmodecoupling}) and Eq.~(\ref{eq:gNLmodecoupling}). 

\section{Inflaton-only case}
\label{sec:inflaton}
In this section we apply the same logic to single-field slow-roll inflation to show that (for perturbations along an attractor) the local curvature perturbations are not modulated by long-wavelength fluctuations, as they are in the curvaton case. We first review the standard calculation of inflationary perturbations in the Hamilton-Jacobi formalism \cite{Salopek:1990jq} (for a recent review see \cite{Hu:2011vr}). We then calculate small-scale perturbations in the presence of a long-wavelength fluctuation in $\delta\phi$ and recover the single-field consistency relation \cite{Maldacena:2002vr,Creminelli:2004yq}.

\subsection{Generation of curvature perturbations in single-field inflation}
In single-field slow-roll inflation the equation of motion for the inflaton is
\be
\label{eq:phieom}
\ddot\phi+3H\dot\phi+\frac{\partial V}{\partial \phi}=0\,.
\ee
In the super-horizon limit $k/a \ll H$, the equation of motion for the perturbations is 
\be
\ddot{\delta\phi} + 3 H\dot{\delta\phi} + \frac{\partial^2 V(\phi)}{\partial \phi^2}\delta\phi = 0\,.
\ee
So the spectrum of inflaton perturbations is given by
\be
\Delta_\phi^2 = \frac{H_{exit}^2}{(2\pi)^2}\,.
\ee
We define the slow-roll parameters as
\be
\epsilon \equiv \frac{3}{2}\frac{\dot\phi^2}{\frac{1}{2}\dot\phi^2+V(\phi)} \quad \, \quad \eta \equiv -\frac{\ddot\phi}{H\dot\phi}\,.
\ee
In the limit that $\epsilon, \eta \ll 1$ the inflaton equation of motion reduces to
\be
\label{eq:attractor}
3H\dot\phi \approx -\frac{\partial V}{\partial \phi}
\ee
which is a first-order equation,  so the value of the field $\phi$ completely specifies the value of $\dot\phi$. 
In this regime we can use the field value $\phi$ as a time variable \footnote{Note, that while Eq.~(\ref{eq:attractor}) is only true to lowest order in $\epsilon$, $\eta$, the full equation of motion Eq.~(\ref{eq:phieom}) can be used to iteratively express $\dot\phi$ and $\ddot\phi$ in terms of $\phi$ to higher order in slow roll parameters. So, the use of $\phi$ as a time variable holds beyond lowest order in the slow roll parameters.}.

The equation of motion for the inflaton can be written as
\be
d\ln \rho_\phi =-2d\ln a\,\epsilon(\phi)
\ee 
If the inflaton is the only source of energy we can relate perturbations in the inflaton energy density to perturbations in the expansion via
\ba
\zeta &=& \delta \ln a \\
&=& -\frac{1}{2\epsilon(\phi)}\frac{\delta\rho_\phi}{\rho_\phi} \\
&=&\frac{\delta\phi}{\sqrt{2\epsilon}M_{pl}}\left(1-\eta/3\right) \rm{sign}(\dot\phi) -\frac{\dot{\delta\phi}}{3\dot\phi}\,,
\ea 
which gives the leading-order in slow-roll parameters expression for the variance of curvature perturbations as
\be
\Delta_\zeta^2 = \left.\frac{H^2_{exit}}{8\pi^2\epsilon(\phi) M_{pl}^2}\right|_{aH = k}\,,
\ee
where $H_{exit}$ again is the value of the Hubble parameter when the mode $k$ crosses the horizon. The spectral index is
\be
(n_s-1)\equiv \frac{d\ln \Delta_\zeta^2}{d\ln k } =2\eta- 4\epsilon + \mathcal{O}(\epsilon^2)\,.
\ee

\subsection{Curvature perturbations on a long-wavelength mode}
We now calculate the amplitude of small-scale curvature perturbations in a region with a longer wavelength fluctuation $\delta\phi_L$. 
The local value of the Hubble parameter in a region with $\delta\phi_L$ is changed to 
\be
\left.H^2 \right|_S = H^2(1+\frac{d\ln H}{d\ln a}\frac{d\ln a}{d\phi}\delta\phi_L)
\ee
and the local slow-roll parameters are 
\ba
\left. \epsilon \right|_S &=&  \epsilon(1+\frac{d\ln \epsilon}{d\ln a}\frac{d\ln a}{d\phi}\delta\phi_L)\,\\
\left. \eta \right|_S &=&  \eta(1+\frac{d\ln \eta}{d\ln a}\frac{d\ln a}{d\phi}\delta\phi_L)\,.
\ea
The mode $k$ now crosses the horizon at 
\be
k =  aH(1+\frac{d\ln a}{d\phi}\delta\phi_L+\frac{d\ln H}{d\ln a}\frac{d\ln a}{d\phi}\delta\phi_L)
\ee
so the small-scale power in the region with a long-wavelength perturbation $\delta\phi_L$ is
\ba
\label{eq:delta2singlefield}
\left. \Delta_\zeta^2\right|_{\delta\phi_L} &=& \Delta^2_\zeta\left(1+\left(2\frac{d\ln H}{d\ln a}-\frac{d\ln \epsilon}{d\ln a}\right.\right.\\
&&\left.\left.-\frac{d\ln\Delta^2}{d\ln k}\left(1+\frac{d\ln H}{d\ln a}\right)\right)  \zeta_L\right)\nn\\
&=&\Delta_\zeta^2 
\ea
where we've identified $d\ln a/d\phi \,\delta\phi_L = \zeta_L$. We see that, in the attractor limit where $\phi$ is the only degree of freedom the value of $\phi$ completely specifies all quantities: the power spectrum of $\delta\phi$, the length scale (set by the horizon), and the amplitude of curvature perturbations. In particular, fluctuations on large scales (or prior fluctuations in $\delta\phi$) do not modulate the locally measured small-scale power. Note that perturbations generated during a non-attractor phase may behave differently (see, e.g. \cite{Namjoo:2012aa,Chen:2013aj,Chen:2013eea}).

To make contact with the Maldacena relation \cite{Maldacena:2002vr}, note that we have used the comoving scale $k$ defined by scaling out by the local observer's scale factor $ae^\zeta_L$. Rewriting Eq.~(\ref{eq:delta2singlefield}) in terms of $k_{global}$ reintroduces the dependence on $\zeta_L$,
\be
\Delta^2(k_{global})  = \Delta^2(k_{local})  - \frac{d\Delta^2}{d\ln k} \zeta_L \,.
\ee
For a detailed discussion of how to relate the consistency relation to observables for an observer with access to both the long- and short- wavelength modes see \cite{Tanaka:2011aj,Creminelli:2011sq,Senatore:2012nq,Pajer:2013ana}.

\section{Discussion}
\label{sec:discussion}

In \S \ref{sec:curvaton} we explicitly computed the dependence of local curvature fluctuations generated by the curvaton on the amplitude of longer-wavelength modes $\zeta_L$. In the case where the curvaton dominates the energy density at decay, a closed-form non-linear expression relates $\zeta$ to the local values of $\delta\sigma$ and $\sigma_*$. From Eq~(\ref{eq:rhophiHubble}), Eq.~(\ref{eq:Deltasigma2}) and Eq.~(\ref{eq:curvnonlinear}) it is clear that the sole effect of a long-wavelength mode $\delta\sigma_L$ is to shift the apparent value of the curvaton field at the end of inflation from $\sigma_*$ to $\sigma_* + \delta\sigma_L$.  In the case where radiation is also present at curvaton decay, the amplitude of curvature perturbations depends on $\delta\sigma$, $\sigma_*$ and $r$, a parameter quantifying the relative amounts of curvaton and radiation at the decay time. A long-wavelength mode $\delta \sigma_L $ may have a larger effect on the local statistics because the value of $r$ at the decay time is shifted in addition to the reference value of $\sigma_*$.  However, a local observer in a region with large $\delta\sigma_L$ would again interpret their local statistics as consistent with the predictions of the curvaton model with a quadratic potential but with a shifted value of the field position at the decay time $\sigma_* \rightarrow \sigma_* + \delta\sigma_L$. 

The curvaton example also gives some insight into the scenarios in \cite{Nelson:2012sb,LoVerde:2013xka} in which the locally observed statistics of $\zeta$ appear only weakly non-Gaussian despite the fact that the global statistics of $\zeta$ are strongly non-Gaussian. Suppose that on average the global value of $\sigma_*$ is   $\lsim \delta\sigma$. In this case a super observer would describe the statistics of $\zeta$ as strongly non-Gaussian (the dimensionless cumulants of $\zeta$ are $\mathcal{O}(1)$). If inflation lasts for a long time there will be some Hubble-sized patches where the curvaton field has experienced many kicks away from the potential minimum giving a local value of $\sigma_*$ that is large compared with the typical $\delta\sigma$. These regions, with $\left.\sigma_*\right|_S \gg \delta\sigma$, will have statistics that appear to be only weakly non-Gaussian. Note that $\sigma_*$ being  large in comparison to $\delta\sigma \sim H$ does not spoil the criterion $\rho_\sigma \ll \rho_\phi$. 

For comparison, we repeated the calculation of the dependence of local curvature perturbations on long-wavelength perturbations in single-field slow-roll inflation. We see that, as expected, for perturbations generated during inflation on an attractor solution the small-scale power spectrum does not vary with $\zeta_L$. 

\noindent{\bf Acknowledgments}\\
The author thanks Christian Byrnes, Kurt Hinterbichler, Matthew Johnson, Michael Salem, Sarah Shandera, and Matias Zaldarriaga. M.L. is especially grateful to Wayne Hu for helpful discussions and asking the questions that initiated these calculations. M.L. is supported by U.S. Dept. of Energy contract DE-FG02-90ER-40560.

\bibliographystyle{ieeetr}
\bibliography{boxes}

\end{document}